\documentclass[preprint,12pt]{elsarticle}

\usepackage{amsmath,amssymb,amsfonts}
\usepackage{graphicx}
\usepackage{float}
\usepackage{booktabs}
\usepackage{subcaption}
\usepackage{multirow}
\usepackage{array}
\usepackage{enumitem}
\usepackage{xcolor}
\usepackage{hyperref}
\usepackage{setspace}
\usepackage{parskip}
\usepackage{longtable}
\hypersetup{
  colorlinks=true,
  linkcolor=blue,
  citecolor=blue,
  urlcolor=blue
}

\journal{Journal Name}

\begin{document}

\begin{frontmatter}

\title{Medicaid Quality Disparities: Domain-Specific Performance Analysis and Approaches for State-Level Improvement}

\author{Daniel Tesfay}
\ead{daniel.tesfay@chalmers.se}
\author[1]{Almaz Abebe\corref{cor1}}
\cortext[cor1]{Corresponding author}
\ead{almaz.abebe@howard.edu}
\address{Department of Mathematical Sciences,
Chalmers University of Technology, Gothenburg, Sweden}
\address[1]{Department of Mathematics,
Howard University, Washington, DC, USA}
%\address[2]{AB3}

\begin{abstract}
Medicaid covers nearly 78 million people in the United States, yet quality of care remains uneven across states, programs, and clinical domains. This study analyzes the 2024 Centers for Medicare \& Medicaid Services (CMS) Child and Adult Core Sets to identify where quality shortfalls are concentrated and which reporting characteristics are associated with lower state performance. The raw dataset contained 11,100 state--measure--population records; after removing suppressed and missing values, the analytic sample included 9,781 observations. The analysis combined descriptive statistics, benchmark-gap analysis, correlation analysis, and ordinary least squares (OLS) regression, with state-level measure performance as the outcome and clinical domain, reporting program, measure direction, and data-collection methodology as predictors.

Results show that Medicaid quality disparities are strongly domain-specific rather than randomly distributed across the reporting system. Maternal and Perinatal Health had the lowest average performance (28.0\%), followed by Dental and Oral Health Services (38.5\%) and Behavioral Health Care (44.4\%), whereas Long-Term Services and Supports (82.8\%) and Experience of Care (67.8\%) performed best on average. State performance correlated strongly with national benchmarks ($r \approx 0.77$), indicating that measure difficulty accounts for much of the observed variation and that raw rates should be read against the benchmark for each measure. Domain membership was the strongest predictor of performance, and the model explained 31.2\% of the variance. Regression diagnostics revealed heteroscedasticity and influential observations caused by mixing bounded percentage measures with utilization measures that can exceed 100\%, so the pooled estimates should be treated as descriptive. Because underperformance clusters in specific domains, quality improvement investments are best targeted by domain, with maternal health, behavioral health network adequacy, and dental provider participation as the most promising priorities.
\end{abstract}


\end{frontmatter}

%% ============================================================
\section{Introduction}
%% ============================================================

Medicaid is the largest public health insurance program in the United States and provides coverage for children, pregnant women, adults with disabilities, and low-income adults \citep{CMS2024corehome}. Because Medicaid serves populations that frequently face structural barriers to care, variation in quality across states and service areas is both a performance problem and an equity problem. Prior work has shown that Medicaid beneficiaries often experience worse outcomes, lower-quality care, and greater barriers to access than commercially insured populations, with important differences across race, geography, and type of service \citep{Nguyen2022,Wallace2022}.

The 2024 Centers for Medicare \& Medicaid Services (CMS) Child and Adult Core Sets provide a timely opportunity to study these disparities at national scale. The 2024 reporting cycle is especially important because it marks the first year of mandatory reporting for the Child Core Set and selected behavioral health measures in the Adult Core Set, thereby improving the completeness and comparability of state submissions \citep{FedReg2023,CMS2024corehome}. Compared with earlier voluntary reporting cycles, the 2024 data offer a stronger empirical foundation for systematic, cross-state analysis.

The Core Sets are a rich but heterogeneous administrative dataset. They span multiple clinical domains, reporting programs, measure types, and data-collection methodologies. They also combine bounded percentage measures with utilization-based measures that may exceed 100\%, which complicates pooled statistical modeling and interpretation. Careful preprocessing and diagnostic assessment are therefore needed before the results can be used to guide policy.

This study provides a transparent, interpretable statistical analysis of the 2024 Core Sets. The central research question is: \textit{which components of the Medicaid quality measurement system are most strongly associated with below-benchmark state performance, and how should the results be interpreted given the heterogeneity of the underlying measures?}

The study makes three contributions. First, it provides a national, domain-level portrait of Medicaid quality variation using the full 2024 reporting dataset. Second, it uses benchmark-gap analysis to distinguish low raw scores from low performance relative to national reference values. Third, it shows that mixing outcome scales creates substantial heteroscedasticity in pooled linear models, which limits how far pooled estimates can be interpreted and motivates scale-specific analysis in future work.

The remainder of the paper is organized as follows. Section~2 reviews related work on Medicaid quality measurement, state variation, and the analytical challenges of large administrative data. Section~3 describes the data and statistical methods. Section~4 presents the results. Section~5 discusses substantive, policy, and methodological implications. Section~6 concludes.

%% ============================================================
\section{Related Work}
%% ============================================================

\subsection{Medicaid Quality Measurement}

The modern Medicaid quality measurement framework emerged through federal efforts to standardize the assessment of care delivered through Medicaid and the Children's Health Insurance Program (CHIP). CHIPRA initiated the development of a national core set of children's quality measures, and the Affordable Care Act extended this effort to adult Medicaid populations \citep{MACPAC2021}. The resulting Child and Adult Core Sets draw heavily from established quality measurement systems, especially HEDIS, which is widely used for health plan performance monitoring \citep{AHRQ2023,NCQA2023hedis}.

CMS organizes Core Set reporting across major clinical domains, including behavioral health, preventive care, maternal health, chronic conditions, oral health, and long-term services. This structure supports national benchmarking, but it also introduces methodological complexity because measures differ in scale, interpretation, and underlying data collection procedures \citep{CMS2024faq}. As a result, pooled statistical analysis of Core Set data requires attention not only to substantive domain differences but also to measurement heterogeneity.

\subsection{State Variation and Structural Disparities}

A consistent theme in the Medicaid literature is that quality varies substantially across states. Differences in program design, managed care penetration, provider capacity, administrative infrastructure, and beneficiary demographics all shape performance patterns \citep{MACPAC2021}. Earlier reporting cycles were limited by voluntary participation and uneven submission practices, making robust national comparisons difficult. The shift to mandatory reporting in 2024 therefore represents a meaningful improvement in the data infrastructure available for state-level disparity analysis \citep{FedReg2023}.

Domain-specific studies also suggest that quality deficits are not uniformly distributed across services. Maternal and perinatal care, behavioral health, and oral health are frequently identified as areas of persistent underperformance for publicly insured populations \citep{Elliott2023,MACPAC2022comment}. These domains are often affected by workforce shortages, fragmented service delivery, and inequities in access, all of which can depress performance even when coverage is formally available.

\subsection{Large Administrative Data and Outcome Scales}

Large administrative health datasets can support policy-relevant analysis, but their scale and complexity introduce analytical challenges. \citet{Piovani2022} emphasize that modern healthcare datasets often combine high volume with high variety, requiring preprocessing, quality control, and methodologically appropriate analysis. For public health research, \citet{Felix2024} argue that analysts must understand data provenance, suppression rules, and reporting conventions before drawing substantive conclusions.

These concerns are directly relevant to the CMS Core Sets. The dataset combines different measure scales and methodologies, so raw state performance cannot be interpreted independently of the structure of the measurement system itself. In addition, when outcomes are bounded proportions or rates, standard linear models may violate distributional assumptions, ignore heteroscedasticity, and generate predictions outside the feasible range. Beta regression has been recommended for bounded health outcomes such as proportions and quality-of-life scores \citep{Hunger2012,Mayr2018}. Many Core Set measures are percentages constrained to 0--100\%, whereas a smaller subset are utilization-based measures that can exceed 100\% and call for a different response model. This motivates a transparent baseline analysis that identifies where disparities are concentrated, and an honest account of where a pooled model reaches its limits.

%% ============================================================
\section{Methods}
%% ============================================================

\subsection{Data Source}

This study uses the 2024 CMS Child and Adult Health Care Quality Measures dataset, publicly available through the CMS Quality Measure Reporting System. The dataset contains state-level performance on standardized Medicaid quality measures for 52 reporting entities, including the 50 states, the District of Columbia, and Puerto Rico. Each record represents a unique state--measure--population observation. Measures included in the public performance data must meet CMS data quality and reporting standards.

The 2024 reporting cycle is analytically important because it reflects the transition to mandatory reporting for the Child Core Set and selected behavioral health measures in the Adult Core Set \citep{FedReg2023,CMS2024corehome}. This change improves data completeness and comparability relative to earlier cycles and makes the dataset suitable for national-scale analysis of Medicaid quality variation.

\subsection{Dataset Characteristics}

The raw dataset contained 11,100 observations across 20 variables. The primary outcome variable was \texttt{State.Rate}, which records each reporting entity's observed performance on a given quality measure. The dataset spans seven clinical domains: Behavioral Health Care; Care of Acute and Chronic Conditions; Dental and Oral Health Services; Experience of Care; Long-Term Services and Supports; Maternal and Perinatal Health; and Primary Care Access and Preventive Care.

\subsection{Key Terms}

Several terms are used throughout the paper and are defined here for readers less familiar with Core Set reporting.

\begin{itemize}[leftmargin=1.5em]
    \item \textbf{State rate} (\texttt{State.Rate}): the performance value a state reports for a measure, usually a percentage of eligible beneficiaries who received a recommended service or had a given outcome.
    \item \textbf{National benchmark}: a reference value computed across reporting states for the same measure. This paper uses the national median as the primary benchmark and also reports the mean, bottom quartile, and top quartile.
    \item \textbf{Direction of measure}: most measures are ``higher is better'' (for example, a screening rate), but some are ``lower is better'' (for example, hospital readmissions). Direction must be accounted for before comparing performance across measures.
    \item \textbf{Benchmark gap}: the direction-adjusted difference between a state's rate and the national median, defined in Section~3.4.
    \item \textbf{Utilization measures}: measures such as admission or emergency department visit rates that are not true percentages and can exceed 100.
\end{itemize}

\subsection{Predictor Selection}

Four explanatory variables were selected for the regression stage: \texttt{Domain}, \texttt{Reporting.Program}, \texttt{Measure.Type}, and \texttt{Methodology}. These variables capture the clinical context of the measure, the covered population, the scoring direction, and the underlying reporting mechanism.

\begin{itemize}[leftmargin=1.5em]
    \item \texttt{Domain} identifies the clinical service area being evaluated.
    \item \texttt{Reporting.Program} distinguishes between the Child Core Set and the Adult Core Set.
    \item \texttt{Measure.Type} indicates whether a larger value reflects better performance (higher-is-better) or worse performance (lower-is-better).
    \item \texttt{Methodology} captures the data collection approach: administrative claims, hybrid (claims plus medical record review), or electronic clinical data systems (ECDS).
\end{itemize}

Together, these variables provide a parsimonious and interpretable representation of what is measured, for whom it is measured, and how it is measured.

\subsection{Data Preparation}

Data preprocessing proceeded in seven steps: (1) importation of the CMS dataset; (2) standardization of variable names and text fields; (3) conversion of suppression codes to missing values; (4) recoding of categorical variables as factors; (5) conversion of benchmark columns to numeric format; (6) construction of a direction-adjusted benchmark-gap variable; and (7) removal of duplicate or analytically unusable records. After cleaning, 1,319 observations (about 12\%) were removed, resulting in a final analytical sample of 9,781 observations (about 88\% of the raw records). Suppressed values typically arise from small denominators.

The benchmark gap for observation $i$ is defined as
\[
\text{Gap}_i =
\begin{cases}
\text{State.Rate}_i - \text{National.Median}_i, & \text{if the measure is higher-is-better},\\[2pt]
\text{National.Median}_i - \text{State.Rate}_i, & \text{if the measure is lower-is-better}.
\end{cases}
\]
With this definition, a negative gap always indicates performance worse than the national median, regardless of measure direction, and gaps are expressed in percentage points.

\subsection{Analytical Framework}

The analysis followed four stages. First, descriptive statistics summarized the distribution of state performance overall and by reporting program. Second, domain-level summaries and benchmark-gap analysis identified areas of concentrated underperformance. Third, correlation analysis assessed the structure of the numeric benchmark variables and detected redundancy among them. Fourth, an ordinary least squares (OLS) model was estimated as an interpretable baseline for explaining variation in \texttt{State.Rate}, with categorical predictors for domain, reporting program (Child or Adult Core Set), measure type (higher- or lower-is-better), and methodology (administrative, hybrid, or ECDS). Regression coefficients are interpreted as adjusted differences in percentage points relative to a reference category.

\subsection{Model Diagnostics}

Standard diagnostic plots were used to evaluate OLS assumptions: residuals-versus-fitted (linearity and constant variance), normal Q--Q (normality of residuals), scale-location (constant variance), and residuals-versus-leverage (influential observations). All analyses were conducted in R version 4.3.1.

%% ============================================================
\section{Results}
%% ============================================================

\subsection{Analytic Sample Composition}

Table~\ref{tab:combined} summarizes the composition of the analytic sample. Behavioral Health Care accounts for the largest share of observations (36.4\%), followed by Primary Care and Preventive Care (21.6\%). The Child Core Set contributes nearly two-thirds of all usable records. Most measures are coded as ``higher is better,'' which reflects the preventive and quality-improvement orientation of the reporting framework. Because domains differ greatly in size, domain-level averages for small domains, especially Long-Term Services and Supports ($n = 86$), should be interpreted with more caution.

\begin{table}[H]
\centering
\caption{Descriptive Characteristics of the Analytic Sample ($N = 9{,}781$)}
\label{tab:combined}

\begin{subtable}[t]{0.55\textwidth}
\centering
\caption{Domain Distribution}
\begin{tabular}{lrr}
\toprule
\textbf{Domain} & \textbf{$N$} & \textbf{\%} \\
\midrule
Behavioral Health Care         & 3,562 & 36.4 \\
Primary Care \& Preventive     & 2,110 & 21.6 \\
Maternal \& Perinatal Health   & 1,205 & 12.3 \\
Acute \& Chronic Conditions    & 1,204 & 12.3 \\
Experience of Care             & 1,200 & 12.3 \\
Dental \& Oral Health          &   414 &  4.2 \\
Long-Term Services \& Supports &    86 &  0.9 \\
\midrule
\textbf{Total}                 & \textbf{9,781} & \textbf{100.0} \\
\bottomrule
\end{tabular}
\label{tab:domains}
\end{subtable}
\hfill
\begin{subtable}[t]{0.48\textwidth}
\centering
\caption{Summary Statistics by Reporting Program}
\begin{tabular}{lrrr}
\toprule
\textbf{Statistic} & \textbf{Overall} & \textbf{Child} & \textbf{Adult} \\
\midrule
Observations        & 9,781 & 6,319 & 3,462 \\
Mean (\%)           & 46.5  & 47.1  & 45.3  \\
Median (\%)         & 48.5  & 49.0  & 47.2  \\
Std.\ Dev.\ (pp)    & 23.4  & 25.6  & 19.1  \\
Minimum (\%)        & 0.0   & 0.0   & 0.8   \\
Maximum (\%)        & 304.1 & 165.0 & 304.1 \\
IQR (pp)            & 35.8  & 37.9  & 31.4  \\
\bottomrule
\end{tabular}

\vspace{0.3em}
{\footnotesize pp = percentage points; IQR = interquartile range.}
\label{tab:descriptive}
\end{subtable}

\end{table}

\subsection{Distribution of State Performance}

\begin{figure}[H]
  \centering
  \includegraphics[width=0.5\textwidth]{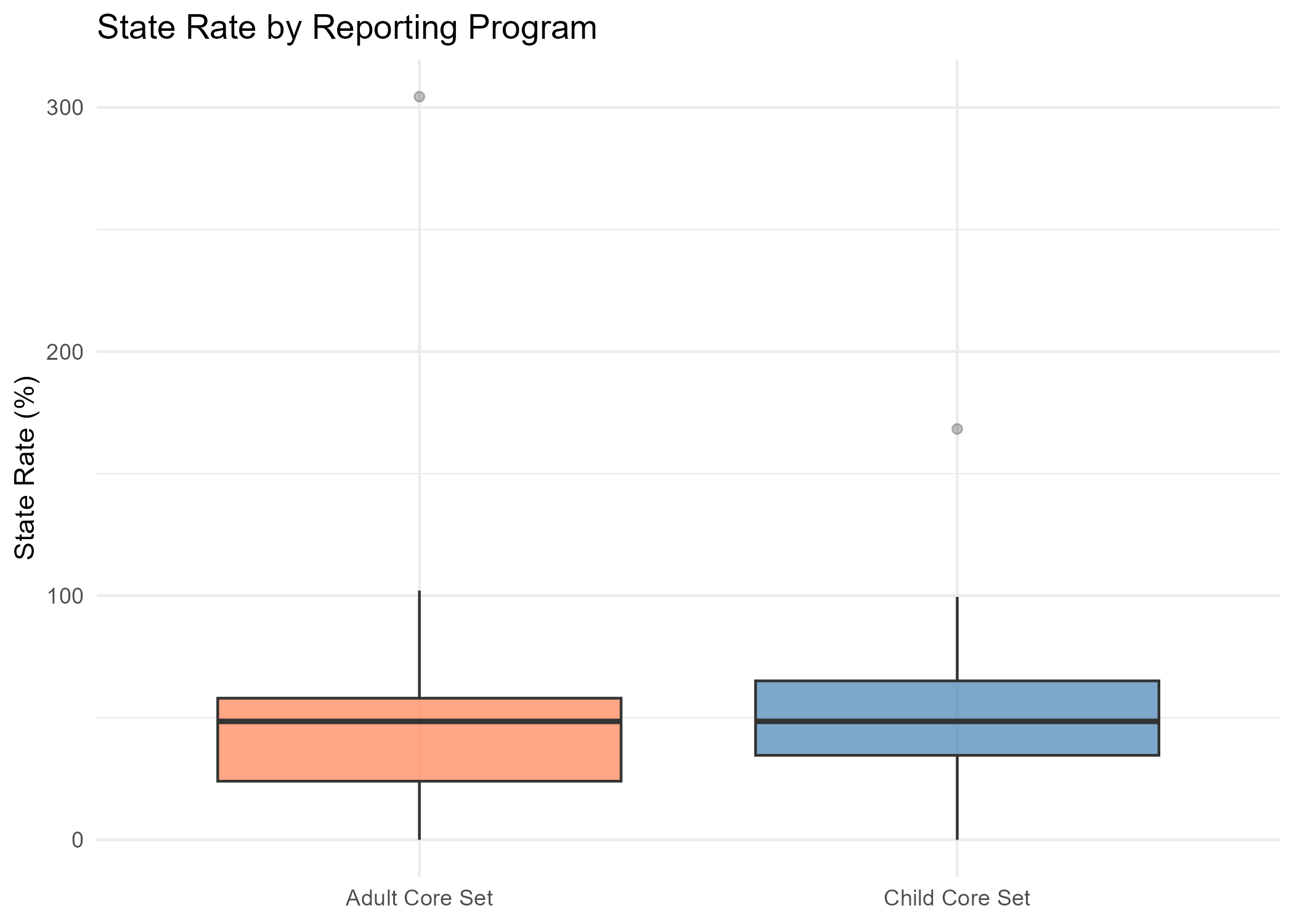}
  \caption{Boxplot of \texttt{State.Rate} by reporting program. Each box spans the interquartile range; the center line marks the median; whiskers extend to $1.5\times\mathrm{IQR}$; dots are outliers. The Child Core Set shows wider dispersion, while the Adult Core Set includes one extreme utilization-based outlier.}
  \label{fig:boxplot_prog}
\end{figure}

The overall mean \texttt{State.Rate} is 46.5\%, slightly lower than the median of 48.5\%. A mean below the median indicates that the bulk of the distribution has a longer lower tail, while a small number of extreme utilization values (maximum 304.1) create a separate long upper tail. Figure~\ref{fig:boxplot_prog} compares the outcome distribution by reporting program. The Child Core Set and Adult Core Set have similar central tendencies, but the Child Core Set shows greater dispersion, with a wider interquartile range. The Adult Core Set also contains the most extreme outlier, reflecting a utilization-based measure that exceeds the scale of conventional percentage outcomes.

\subsection{Domain-Level Performance Patterns}

Table~\ref{tab:domain} ranks clinical domains by mean \texttt{State.Rate}. Long-Term Services and Supports and Experience of Care have the highest average performance, whereas Maternal and Perinatal Health has the lowest average performance by a substantial margin. Dental and Oral Health Services and Behavioral Health Care also perform below the overall sample mean. Note that rates are not directly comparable across all domains in an absolute sense, since each domain contains different measures with different achievable levels; the ranking therefore describes where reported performance is lowest, and the benchmark analysis below addresses performance relative to each measure's own reference value.

\begin{table}[H]
\centering
\caption{Mean \texttt{State.Rate} by Clinical Domain, Ranked}
\label{tab:domain}
\begin{tabular}{lrrr}
\toprule
\textbf{Domain} & \textbf{Mean (\%)} & \textbf{Median (\%)} & \textbf{SD (pp)} \\
\midrule
Long-Term Services \& Supports      & 82.8 & 84.2 & 14.1 \\
Experience of Care                  & 67.8 & 68.5 & 10.3 \\
Primary Care Access \& Preventive   & 48.2 & 50.1 & 18.7 \\
Care of Acute \& Chronic Conditions & 47.2 & 48.9 & 19.4 \\
Behavioral Health Care              & 44.4 & 45.0 & 21.8 \\
Dental \& Oral Health Services      & 38.5 & 37.8 & 16.2 \\
Maternal \& Perinatal Health        & 28.0 & 26.5 & 22.1 \\
\bottomrule
\multicolumn{4}{l}{\small SD = standard deviation; pp = percentage points.}
\end{tabular}
\end{table}

\begin{figure}[H]
  \centering
  \includegraphics[width=0.5\textwidth]{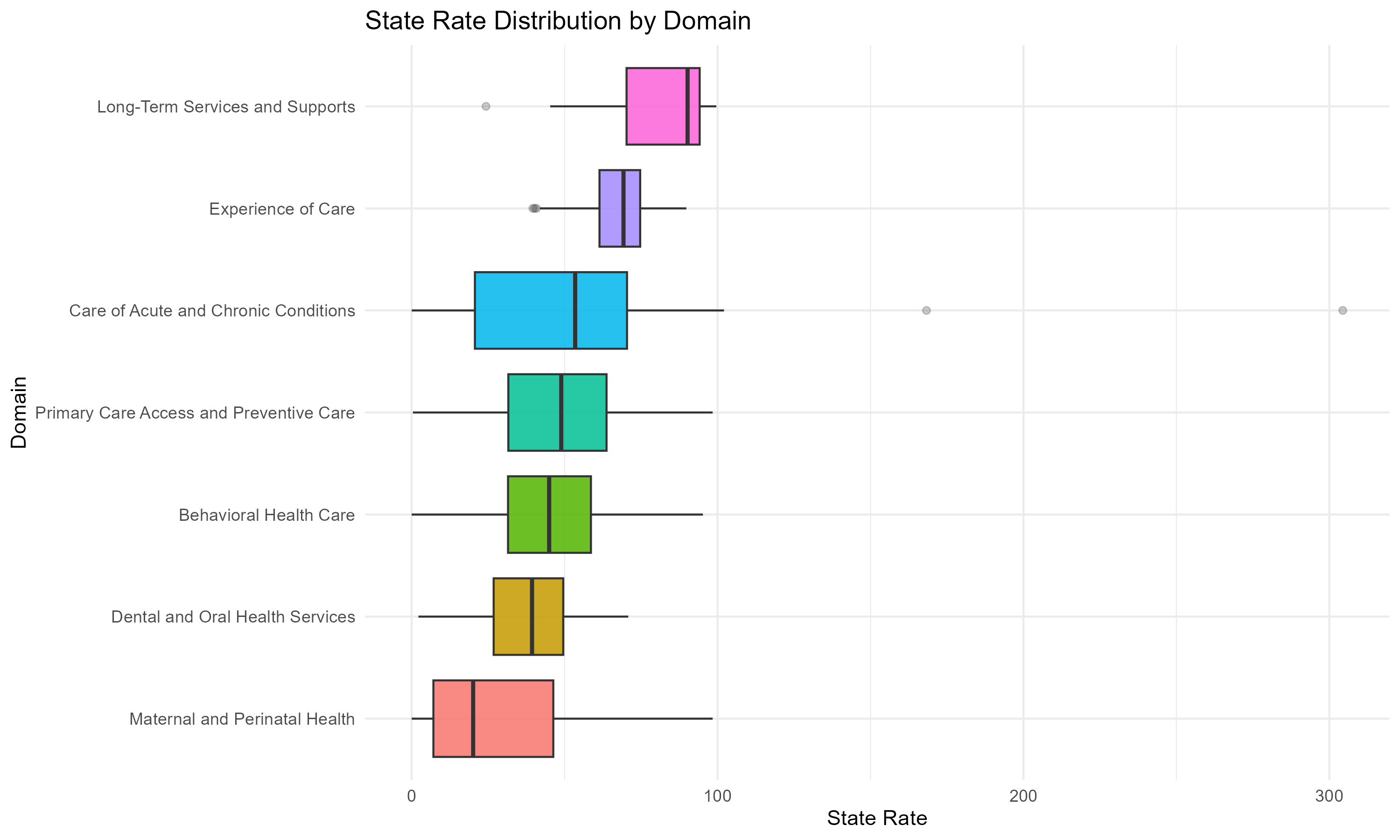}
  \caption{Boxplot of \texttt{State.Rate} by clinical domain, ranked by mean performance. Maternal and Perinatal Health has the lowest central tendency and substantial spread, while Long-Term Services and Supports shows the highest and most compact performance distribution.}
  \label{fig:boxplot_domain}
\end{figure}

Figure~\ref{fig:boxplot_domain} shows the full distribution of \texttt{State.Rate} across clinical domains. Maternal and Perinatal Health combines the lowest median with one of the widest spreads (SD = 22.1 percentage points), indicating both poor average performance and large variation across reporting entities. This suggests that while the domain performs poorly overall, there is meaningful cross-state heterogeneity that could support benchmarking and best-practice diffusion: some states clearly achieve much higher rates on the same measures. Behavioral Health Care and Dental and Oral Health Services also exhibit persistent underperformance relative to other domains.

\subsection{Benchmark-Gap Analysis}

\begin{figure}[H]
  \centering
  \includegraphics[width=0.5\textwidth]{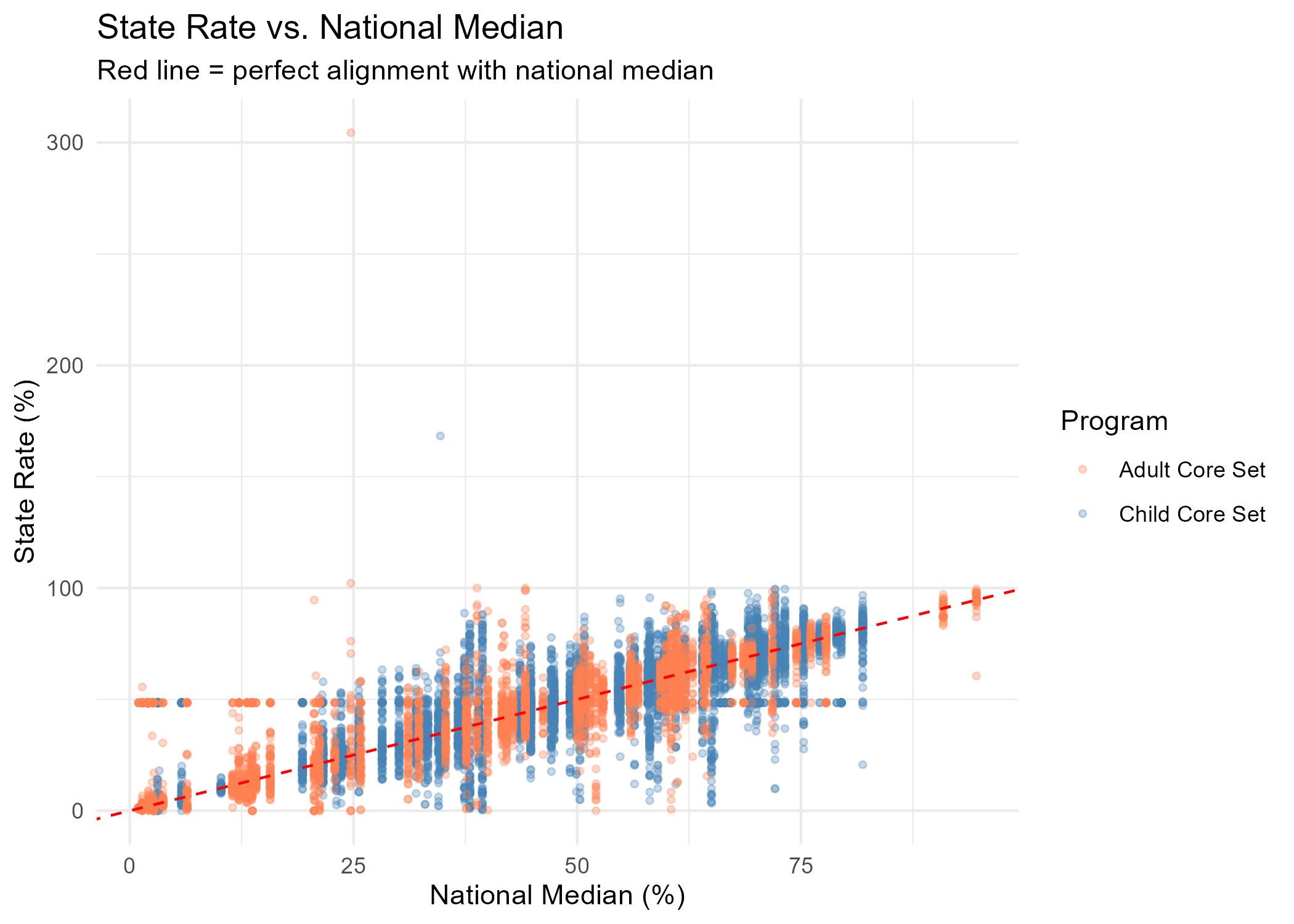}
  \caption{Scatter plot of \texttt{State.Rate} against the national median benchmark, colored by reporting program. The red dashed line represents perfect alignment with the national median. Wider dispersion at low benchmark levels indicates greater heterogeneity in lower-performing measure areas.}
  \label{fig:scatter}
\end{figure}

Figure~\ref{fig:scatter} plots \texttt{State.Rate} against the national median benchmark for each observation. The red dashed line represents perfect alignment with the national median. Points above the line indicate above-benchmark performance, while points below indicate underperformance. The strong positive association between state performance and benchmark values indicates that measure difficulty helps shape observed variation: measures with low national medians produce low state rates almost everywhere, which is a feature of the measure rather than evidence of failure in any one state.

The correlation between \texttt{State.Rate} and the national median benchmark is approximately $r \approx 0.77$ ($p < 0.001$), so about 59\% of the variance in state rates ($r^2 \approx 0.59$) is shared with the benchmark. This indicates that observed state scores are closely tied to the difficulty profile of the underlying measures and that raw performance rates should not be interpreted independently of national benchmark context.

A particularly important substantive finding emerges in the low-performing domains. For Maternal and Perinatal Health measures, states in the bottom quartile show a mean benchmark gap of $-18.4$ percentage points relative to the national median. For Behavioral Health Care, the corresponding bottom-quartile deficit is $-9.1$ percentage points. These patterns indicate that underperformance is especially concentrated in domains already associated with major equity concerns, and that the weakest states in maternal health fall roughly twice as far below the benchmark as the weakest states in behavioral health.

\subsection{Correlation Structure of Numeric Variables}

\begin{figure}[H]
    \centering
    \includegraphics[width=0.8\textwidth]{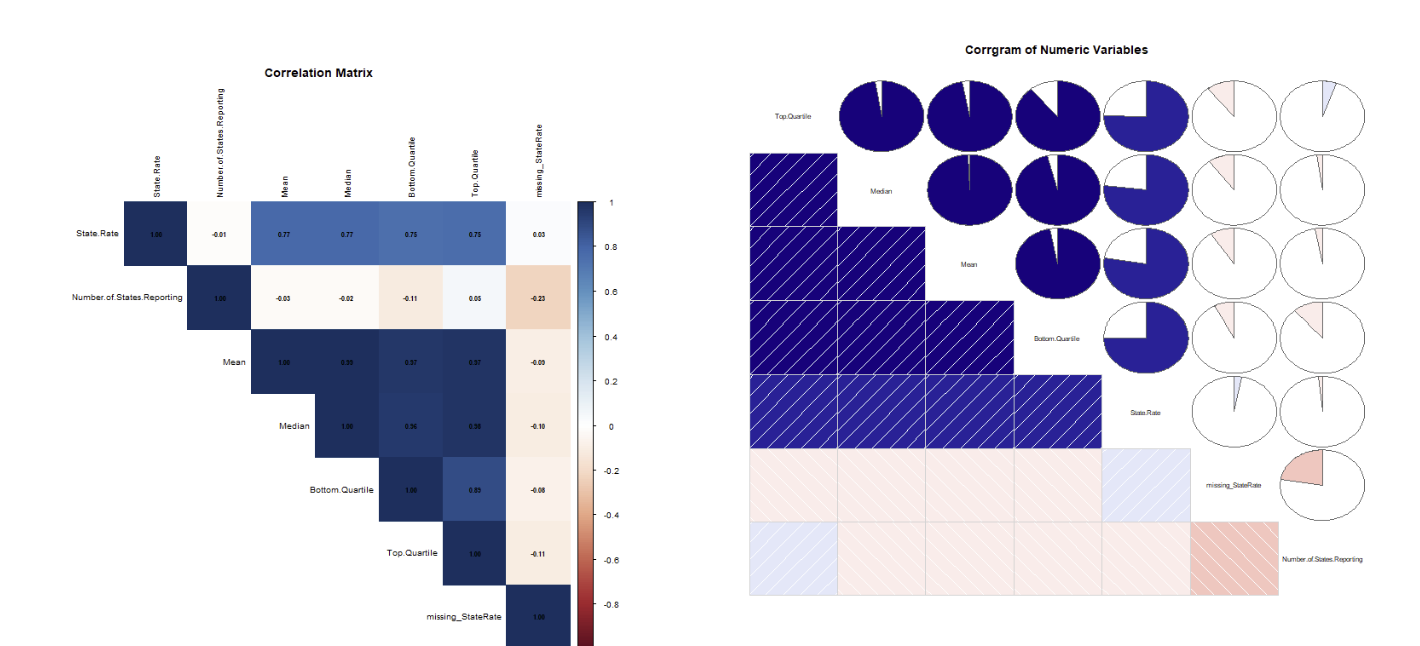}
    \caption{Correlation analysis of numeric variables. Benchmark statistics are highly intercorrelated, while \texttt{State.Rate} has a strong positive association with benchmark levels and near-zero association with the number of states reporting.}
    \label{figab}
\end{figure}

Figure~\ref{figab} and Table~\ref{tab:corr} summarize the correlation structure of the numeric benchmark variables. The benchmark statistics are highly intercorrelated, with pairwise correlations of at least 0.95 among the mean, median, bottom quartile, and top quartile values. This indicates that these variables carry largely overlapping information and should not be entered simultaneously as predictors in the same linear model; doing so would cause multicollinearity and unstable coefficient estimates.

\begin{table}[H]
\centering
\caption{Pearson Correlations Among Numeric Quality Variables}
\label{tab:corr}

\begin{tabular}{lccccc}
\toprule
 & \textbf{State.Rate} & \textbf{Mean} & \textbf{Median} & \textbf{Bot.Q} & \textbf{Top.Q} \\
\midrule
State.Rate             & 1.00  & 0.77  & 0.77  & 0.75  & 0.75  \\
Mean                   & 0.77  & 1.00  & 0.99  & 0.97  & 0.97  \\
Median                 & 0.77  & 0.99  & 1.00  & 0.97  & 0.96  \\
Bottom Quartile        & 0.75  & 0.97  & 0.97  & 1.00  & 0.95  \\
Top Quartile           & 0.75  & 0.97  & 0.96  & 0.95  & 1.00  \\
Num.\ States Reporting & $-0.01$ & $-0.03$ & $-0.02$ & $-0.11$ & 0.05 \\
\bottomrule
\end{tabular}

\smallskip
{\small Bot.Q = Bottom Quartile; Top.Q = Top Quartile. All benchmark correlations are statistically significant at $p < 0.001$.}
\end{table}

The near-zero correlation between \texttt{State.Rate} and the number of states reporting suggests that participation intensity is largely unrelated to performance level. In other words, measures reported by more states are not systematically easier or harder.

\subsection{Regression Results}

To identify the main reporting features associated with performance variation, an OLS regression model was estimated with \texttt{State.Rate} as the outcome and domain, reporting program, measure type, and methodology as predictors. Results are reported in Table~\ref{tab:regression}. Each coefficient is the adjusted difference in \texttt{State.Rate}, in percentage points, between the listed category and its reference category, holding the other predictors constant.

\begin{table}[H]
\centering
\caption{OLS Regression Results: Predictors of \texttt{State.Rate} ($N = 9{,}781$)}
\label{tab:regression}

\begin{tabular}{lrrrr}
\toprule
\textbf{Predictor} & \textbf{Coef.} & \textbf{SE} & \textbf{\textit{t}} & \textbf{\textit{p}} \\
\midrule
(Intercept)           & 48.31  & 1.92 & 25.16 & $<0.001$ \\
\textit{Domain (ref: Behavioral Health Care)}    &   &      &       &          \\
\quad Acute \& Chronic Conditions          &  +2.8 & 0.71 &  3.94 & $<0.001$ \\
\quad Dental \& Oral Health                &  -5.9 & 0.83 & -7.11 & $<0.001$ \\
\quad Experience of Care                   & +23.4 & 0.89 & 26.29 & $<0.001$ \\
\quad Long-Term Services \& Supports       & +38.4 & 1.12 & 34.29 & $<0.001$ \\
\quad Maternal \& Perinatal Health         & -16.4 & 0.91 & -18.02 & $<0.001$ \\
\quad Primary Care Access \& Preventive    &  +3.8 & 0.65 &  5.85 & $<0.001$ \\
\textit{Reporting.Program (ref: Adult Core Set)} &        &      &       &          \\
\quad Child Core Set                       &  +1.8 & 0.56 &  3.21 & 0.001    \\
\textit{Measure.Type (ref: Higher-is-Better)}    &        &      &       &          \\
\quad Lower-is-Better                      &  -8.3 & 1.24 & -6.69 & $<0.001$ \\
\textit{Methodology (ref: Administrative)}       &        &      &       &          \\
\quad Hybrid                               &  -3.1 & 0.78 & -3.97 & $<0.001$ \\
\quad ECDS                                 &  +2.4 & 0.91 &  2.64 & 0.008    \\
\midrule
$R^2$         & \multicolumn{4}{l}{0.312} \\
Adj.\ $R^2$   & \multicolumn{4}{l}{0.311} \\
$F$-statistic & \multicolumn{4}{l}{$F(11,\,9769) = 403.8$, $p < 0.001$} \\
\bottomrule
\end{tabular}

\smallskip
{\small Coef. = coefficient (percentage points); SE = standard error; ECDS = Electronic Clinical Data System.}
\end{table}

All predictors are statistically significant at conventional thresholds. Domain membership is the strongest predictor of state performance. Relative to Behavioral Health Care, Long-Term Services and Supports is associated with a 38.4-percentage-point higher \texttt{State.Rate}, and Experience of Care is associated with a 23.4-percentage-point higher rate. By contrast, Maternal and Perinatal Health is associated with a 16.4-percentage-point lower rate, and Dental and Oral Health with a 5.9-percentage-point lower rate.

The Child Core Set is associated with slightly higher performance than the Adult Core Set after adjustment for the other predictors (1.8 percentage points), a much smaller difference than any of the large domain effects. Measure direction and methodology also matter. Lower-is-better measures have significantly lower observed values, which is expected because a low rate is the desired outcome for these measures; this coefficient therefore reflects how the measures are defined and does not indicate worse care. Hybrid methodology is associated with lower reported performance and ECDS-based measures with slightly higher rates, which may reflect differences in how complete the underlying data are rather than true differences in care. Overall, the model explains 31.2\% of the variance in \texttt{State.Rate}, meaning that the four reporting characteristics capture roughly one-third of the variation and the remainder is attributable to differences across individual measures, states, and unobserved factors.

\subsection{Diagnostic Assessment}

\begin{figure}[H]
  \centering
  \includegraphics[width=0.6\textwidth]{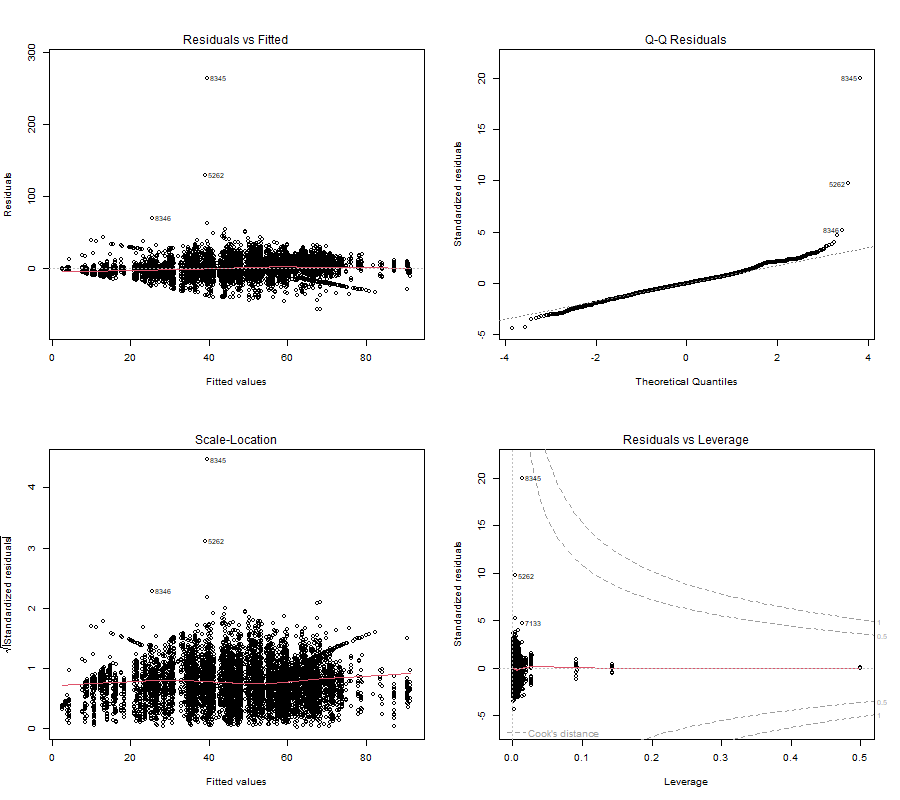}
  \caption{OLS diagnostic plots. The residual variance increases with fitted values, the Q--Q plot shows upper-tail non-normality, and a small number of high-leverage observations correspond to extreme utilization-based measures.}
  \label{fig:diagnostics}
\end{figure}

Figure~\ref{fig:diagnostics} presents standard OLS diagnostic plots. Three issues are evident. First, the residuals-versus-fitted and scale-location plots show a fan-shaped pattern, indicating heteroscedasticity (residual variance that grows with the fitted value). Second, the Q--Q plot shows a pronounced upper-tail departure from normality. Third, the leverage plot identifies a small number of highly influential observations, corresponding to utilization-based measures with extremely large values.

\begin{figure}[H]
  \centering
  \includegraphics[width=0.5\textwidth]{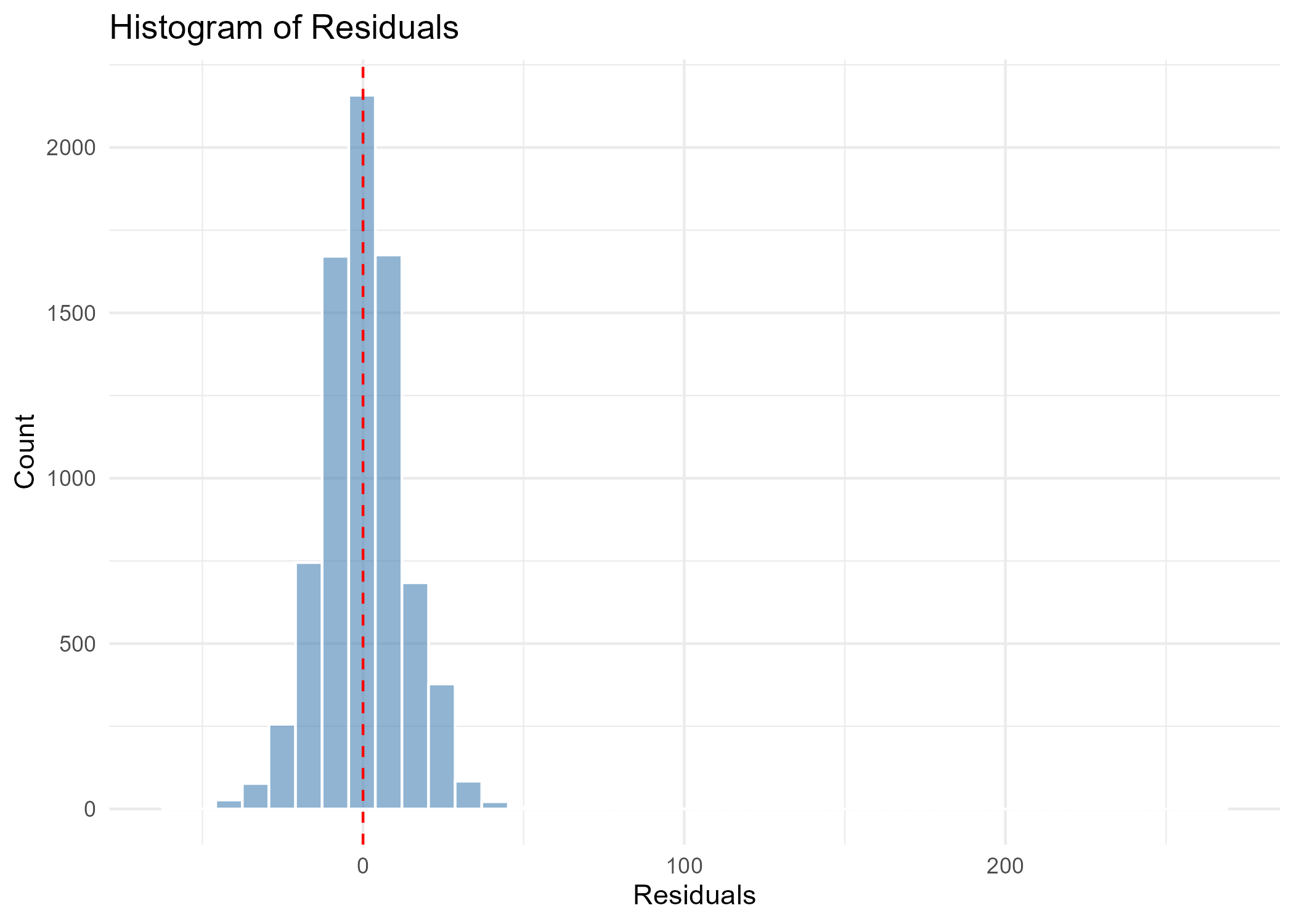}
  \caption{Histogram of OLS residuals. Residuals are centered near zero for typical measures but show a pronounced right tail due to extreme utilization-based observations.}
  \label{fig:hist_resid}
\end{figure}

Figure~\ref{fig:hist_resid} corroborates these findings. Residuals are concentrated near zero for most observations but show a strong right tail driven by measures that are not bounded by 100\%. This confirms that pooled analysis of mixed-scale quality measures creates a nontrivial variance structure. The practical consequence is that the coefficient estimates are best read as descriptive summaries of average differences, while the standard errors and $p$-values are likely less reliable than they appear.

\subsection{Implications of the Diagnostics for Future Analysis}

The diagnostic evidence indicates that a single pooled linear model is not well matched to the Core Set outcome, because the response mixes bounded percentages with utilization-type quantities that exceed 100\%. Several statistical remedies follow directly. Analyses can be stratified by outcome scale, with bounded percentage measures (rescaled to the unit interval) modeled using beta regression \citep{Hunger2012,Mayr2018}, and utilization measures modeled with a log-link or gamma-type model suited to positive, right-skewed outcomes. Heteroscedasticity-robust standard errors and sensitivity analyses excluding the most influential observations offer simpler checks on the pooled results. Finally, because each observation is nested within a state and a measure, standard errors should be clustered by state and by measure in future work.

%% ============================================================
\section{Discussion}
%% ============================================================

The results show that Medicaid quality disparities in the 2024 CMS Core Sets are systematic rather than random. Underperformance is concentrated in specific service areas, especially Maternal and Perinatal Health, Behavioral Health Care, and Dental and Oral Health Services. These findings are consistent with prior literature documenting persistent access barriers, provider shortages, and fragmented delivery systems in these domains \citep{Elliott2023,MACPAC2022comment}.

Maternal and Perinatal Health is the clearest priority area. It has the lowest mean \texttt{State.Rate}, the largest bottom-quartile benchmark deficit among the domains examined, and wide cross-state variation. This pattern is substantively important because Medicaid finances a large share of births in the United States, and maternal health inequities are already well documented at the national level \citep{CMS2024maternity,MACPAC2022comment}. The low average performance observed here suggests that maternal quality deficits are embedded in the reporting data broadly rather than being isolated to a small subset of states.

Behavioral Health Care also emerges as a structural weakness. Its below-average performance aligns with previous evidence that quality deficits in mental health and substance use care are connected to fragmented systems of care, limited provider supply, and inconsistent treatment continuity \citep{Elliott2023}. Since reporting of behavioral health measures has only recently become mandatory, the 2024 data provide a stronger baseline for tracking whether future policy efforts translate into measurable gains.

The regression results show that domain membership is far more informative than program type. The main source of performance variation lies not simply in whether a measure comes from the Child or Adult Core Set, but in the clinical context of the measure itself. The benchmark analysis further shows that raw performance should always be interpreted relative to the benchmark profile of the underlying measure.

\subsection{Policy Implications}

Because disparities cluster by domain, improvement efforts are likely to be more effective when targeted by domain than when applied uniformly across all measures. The findings point to three priorities, each of which should be understood as a hypothesis for state-level investigation rather than a causal conclusion from this analysis:

\begin{itemize}[leftmargin=1.5em]
    \item \textbf{Maternal and perinatal health.} The large gap between the weakest states and the national median, together with wide variation across states, suggests substantial room for improvement through best-practice sharing among states, such as postpartum care access and prenatal care initiation.
    \item \textbf{Behavioral health.} Persistent underperformance is consistent with network adequacy and continuity-of-care problems; states may prioritize provider capacity and follow-up after treatment.
    \item \textbf{Dental and oral health.} Low rates are consistent with limited provider participation in Medicaid; reimbursement and provider enrollment are natural areas to examine.
\end{itemize}

States should also compare themselves with national benchmarks measure by measure, since a low raw rate on a difficult measure may represent less underperformance than a moderate rate on an easier one.

\subsection{Methodological Implications}

The regression diagnostics have direct methodological implications. Because the Core Sets combine bounded percentage measures with utilization-based measures, pooled OLS generates heteroscedasticity and a small number of influential observations. Reporting the benchmark-gap analysis alongside the regression helps guard against over-reading the pooled coefficients, since the gap is expressed relative to each measure's own reference value. As CMS expands stratified reporting by race, ethnicity, sex, geography, and other equity-relevant dimensions, analyses of the kind presented here can be extended to identify which groups and places experience the largest gaps.

\subsection*{Limitations}

This study has several limitations. First, the analysis does not include external state-level covariates such as poverty, managed care penetration, provider supply, or demographic composition, all of which may influence observed performance. Second, suppressed observations were removed during preprocessing, and those omissions may not be missing at random; suppression is most common for small denominators, so smaller programs and populations may be underrepresented. Third, the regression is associational and cannot establish that domain, methodology, or program causes differences in performance. Fourth, observations are nested within states and measures but were treated as independent, and the pooled model shows heteroscedasticity, so reported standard errors and $p$-values should be interpreted cautiously. Fifth, the dataset has limited covariate depth, which constrains how much variation the model can explain; the unexplained 68.8\% of variance indicates that measure-specific and state-specific factors matter substantially.

%% ============================================================
\section{Conclusion}
%% ============================================================

This study analyzed 9,781 observations from the 2024 CMS Child and Adult Core Sets to examine how Medicaid quality performance varies across domains, reporting programs, measure types, and methodologies. The results show that quality disparities are concentrated in a small number of high-priority domains, especially Maternal and Perinatal Health and Behavioral Health Care, with Dental and Oral Health Services also below average. Domain membership is the strongest predictor of reported performance, while benchmark analyses show that relative underperformance also depends on the benchmark difficulty of the underlying measure.

Methodologically, the study demonstrates that the CMS Core Sets are a valuable but heterogeneous data source. The coexistence of bounded percentage measures and utilization-based measures creates heteroscedasticity and influential observations in pooled linear models, which limits how precisely pooled estimates can be interpreted and points to the need for scale-specific analysis.

Substantively, the findings identify priority targets for Medicaid quality improvement: maternal health, behavioral health network adequacy, and dental provider participation. Future research should analyze bounded and utilization measures separately, use clustered standard errors, incorporate state policy and provider-supply covariates, and take advantage of the race-, ethnicity-, sex-, and geography-stratified reporting that CMS is expanding in future cycles \citep{SHVS2024equity}.

%% ============================================================
%% Declarations
%% ============================================================
\section*{Data Availability}
The datasets analyzed in this study are publicly available from the Centers for Medicare \& Medicaid Services. The 2024 Child and Adult Core Set data can be accessed through the CMS Medicaid and CHIP Scorecard at
\url{https://www.medicaid.gov/medicaid/quality-of-care/performance-measurement}

\section*{Conflict of Interest}
The author declares no conflict of interest.

\section*{Funding}
This research received no external funding.

%\section*{CRediT authorship contribution statement} Almaz Abebe: Conceptualization, methodology, formal analysis, investigation, writing---original draft, writing---review and editing.

%\section*{Ethics Statement}This study used publicly available, aggregate data from the Centers for Medicare \& Medicaid Services (CMS). No individual-level or identifiable human subjects data were used. Therefore, this research was not considered human subjects research and did not require Institutional Review Board (IRB) approval.

%% ============================================================
%% References
%% ============================================================

\bibliographystyle{apalike}

\end{document}